\documentclass[twocolumn,preprintnumbers, amssymb,amsmath,aps,floatfix,prc,nofootinbib,superscriptaddress,showpacs]{revtex4-1}

\usepackage{graphicx}
\usepackage{amsmath,amsthm,amssymb}

\usepackage{color}
\usepackage{bm}
\usepackage{lipsum}
\usepackage{ulem}
\usepackage[bookmarks,
                bookmarksopen = true,
                bookmarksnumbered = true,
                linktocpage,
                colorlinks = true,
                linkcolor = blue,
                urlcolor  = blue,
                citecolor = blue,
                anchorcolor = green,
                hyperindex = true,
                hyperfigures]
                {hyperref}

\begin{document}

\title{Nuclear modification of jet shape for inclusive jets and $\gamma$-jets at the LHC energies}

\author{Ning-Bo Chang}
\email{changnb@mail.ccnu.edu.cn}
\affiliation{Institute of Theoretical Physics, Xinyang Normal University, Xinyang, Henan 464000, China}
\affiliation{Institute of Particle Physics and Key Laboratory of Quark and Lepton Physics (MOE), Central China Normal University, Wuhan, 430079, China}

\author{Yasuki Tachibana}
\email{yasuki.tachibana@mail.ccnu.edu.cn}
\affiliation{Department of Physics and Astronomy, Wayne State University, Detroit, Michigan 48201, USA}

\author{Guang-You Qin}
\email{guangyou.qin@mail.ccnu.edu.cn}
\affiliation{Institute of Particle Physics and Key Laboratory of Quark and Lepton Physics (MOE), Central China Normal University, Wuhan, 430079, China}

\date{\today}

\begin{abstract}

With our coupled jet-fluid model, we study the nuclear modifications of full jets and jet structures for single inclusive jets and $\gamma$-jets in Pb+Pb collisions at $5.02$~ATeV and $2.76$~ATeV.
The in-medium evolution of full jet shower is described by a set of coupled transport equations including the effects of collisional energy loss, transverse momentum broadening and medium-induced splitting process.
The dynamical evolution of bulk medium is simulated by solving relativistic hydrodynamic equation with source term which accounts for the energy and momentum deposited by hard jet shower to soft medium.
Our study demonstrates that the hydrodynamic medium response to jet propagation significantly enhances the broadening of jet shape at large angles and is essential for the cone-size dependence of jet energy loss and nuclear modification factor of inclusive jet production.
It is also found that the nuclear modification pattern of jet shape is sensitive to jet energy but has weak dependence on the flavor of the parton that initiates the jet.
Our result can naturally explain the different nuclear modification patterns of jet shape functions for single inclusive jet and $\gamma$-jet events as observed by the CMS Collaboration, and can be tested in the future by measuring the jet shape function over a wider range of jet energies in heavy-ion collisions.

\end{abstract}

\maketitle

\section{Introduction}
\label{sec:introduction}

In ultra-relativistic heavy-ion collisions, large transverse momentum partons that are produced at early stage of the collisions propagate through the quark-gluon plasma (QGP) and experience elastic scatterings and inelastic radiative processes during their interactions with the medium constituents.
The phenomena involving jet-medium interactions and parton energy loss are usually called jet quenching, and have provided unique opportunities to probe the novel properties of the QGP \cite{Wang:1991xy, Qin:2015srf}.
One important quantity in jet quenching studies is the so-called jet quenching parameter $\hat{q}$ which describes the transverse momentum squared exchanged between the propagating jet partons and the QGP medium per unit length and is directly related to the gluon density of the traversed QGP \cite{Baier:1996sk}. The quenching parameter $\hat{q}/T^3$, with $T$ being the temperature of the medium, is also closely related to the kinetic transport properties of the QGP, such as the shear viscosity to entropy ratio $\eta/s$ \cite{Majumder:2007zh}.

Motivated by early jet quenching measurements at the Relativistic Heavy Ion Collider (RHIC), many studies have focused on the suppression of single inclusive hadron spectra at high transverse momentum, which tends to be sensitive to the energy loss of the leading parton in the jet \cite{Khachatryan:2016odn, Acharya:2018qsh, Aad:2015wga, Burke:2013yra, Xu:2014tda, Chien:2015vja, Andres:2016iys, Cao:2017hhk, Zigic:2018ovr, Xing:2019xae}.
In addition, jet-related correlation measurements such as dihadron and $\gamma$-hadron correlations, have provided additional information on jet-medium interaction, such as jet energy loss and medium-induced transverse momentum broadening effects \cite{Qin:2009bk, Chen:2016vem, Chen:2016cof, Chen:2017zte, Luo:2018pto, Zhang:2018urd, Kang:2018wrs}.
Recently, experimental developments at both RHIC and the Large Hadron Collider (LHC) made it possible to measure the spectra and detailed inner structure of fully reconstructed jets \cite{Aad:2010bu,Chatrchyan:2011sx,Chatrchyan:2012gt,Aad:2014bxa,Chatrchyan:2012gw, Chatrchyan:2013kwa, Aad:2014wha}, which have provided much more detailed information on jet quenching and jet-medium interaction.
The measurements of abundant jet observables in ultra-relativistic heavy-ion collisions have encouraged and stimulated many theoretical studies which aim to understand various details of the interaction between full jets and the QGP medium \cite{Vitev:2009rd,Qin:2010mn,CasalderreySolana:2010eh,Lokhtin:2011qq,Young:2011qx,He:2011pd,MehtarTani:2011tz,Renk:2012cx,Zapp:2012ak,Apolinario:2012cg,Dai:2012am,Qin:2012gp,CasalderreySolana:2012ef,Zapp:2012ak,Majumder:2013re,Majumder:2013re,Ma:2013pha,Senzel:2013dta,Blaizot:2013hx,Wang:2013cia,Fister:2014zxa,Casalderrey-Solana:2014bpa,He:2015pra,Chien:2015hda,Milhano:2015mng,Chang:2016gjp,Tachibana:2017syd,Chien:2016led,Chang:2017gkt,Cao:2017zih,Cao:2017qpx,He:2018xjv}.

Many studies have found that the leading parton energy loss is mainly driven by the inelastic radiative processes \cite{Wicks:2005gt,Qin:2007rn,Schenke:2009ik}, and elastic collisions are usually considered as complementary contribution (note that elastic collisions become more important for heavy quarks with low transverse momentum due to their finite mass \cite{Cao:2013ita, Cao:2015hia, Xing:2019xae}).
However, the nuclear modification of full jet requires a more comprehensive understanding of jet shower evolution in the dynamically evolving QGP medium.
Jets consist of many soft radiated partons as well as the leading partons.
It is interesting to find that the collisional energy loss and medium absorption of soft radiated partons play a crucial role in the modification of jet spectra and substructures \cite{Qin:2010mn,Chang:2016gjp}.
In addition, some jet energy is deposited into the medium via scatterings and absorption, and then propagates as collective flow excitations in the medium.
The jet-induced flows will enhance the hadron emission from the medium around the jet axis direction.
Since these enhanced hadrons are correlated with the jets, they are always measured together with the jets and not subtracted as background.
It has been found that the jet-induced flows can cause significant modifications of the outer soft part of the jets \cite{Tachibana:2017syd,Chen:2017zte,KunnawalkamElayavalli:2017hxo}.

One recent important measurement by the CMS Collaboration is the jet shape function of $\gamma$-jet events in Pb-Pb collisions at the LHC, which shows quite different modification pattern from that of single inclusive jets. For inclusive jets, a clear collimation of energy towards the jet axis is observed together with the broadening effect, while for $\gamma$-jets, only monotonic broadening effect is observed for the jet shape function \cite{Chatrchyan:2013kwa, Sirunyan:2018ncy}.
This difference is sometimes attributed to different flavor compositions of $\gamma$-jets and single inclusive jets: $\gamma$-jets are dominated by quark-initiated jets while the inclusive jets have a significant fraction of gluon-initiated jets \cite{Chien:2015hda}.
While this argument seems to be plausible, other explanations are also in debate.

The purpose of this work is to investigate such difference by performing a detailed study of the flavor and energy dependence of jet shape function in heavy-ion collisions.
In particular, we use our coupled jet-fluid model \cite{Qin:2010mn,Chang:2016gjp,Tachibana:2017syd} to study the full jet modification for single inclusive jets and $\gamma$-jets in Pb+Pb collisions at 2.76 ATeV and 5.02 ATeV.
We will present numerical results for the suppression of inclusive jet spectra, the modification of
$\gamma$-jet momentum imbalance distribution, and the nuclear modification of jet shape for single inclusive jets and $\gamma$-jets. It is found that the hydrodynamic response to the energy and momentum deposited by jets is important to
the cone-size dependence of full jet suppression and jet shape modification.
Our study also indicates that the different jet shape modification patterns for single inclusive jets and $\gamma$-jets observed by CMS Collaboration can be naturally explained by different jet energies in two different measurements.
Our predictions can be tested by future measurement of single inclusive jets with lower jet energies.

The paper is organized as follows.
In Sec.~\ref{sec:framework}, we introduce our coupled jet-fluid model that we use to calculate the evolutions of parton shower and jet-induced flow in the expanding QGP fluid.
In Sec.~\ref{results}, we present and discuss the results on various full jet observables for single inclusive jets and $\gamma$-jets in Pb+Pb collisions at 5.02~ATeV and 2.76~ATeV.
The summary is given in Sec.~\ref{sec:summary}.

\section{Framework}
\label{sec:framework}

In our coupled jet-fluid model~\cite{Qin:2010mn,Chang:2016gjp,Tachibana:2017syd}, the jet shower evolution in the QGP medium is described by solving a set of coupled differential equations for the three-dimensional momentum distributions of quarks and gluons in the jet shower, $f_i(\omega_i, k_{i\perp}^2)=dN_i/d\omega_i dk_{i\perp}^2$, where $i$ denotes the parton species (gluon, or quark plus anti-quark), $\omega_i$ is the parton's energy and $k_{i\perp}$ the parton's transverse momentum with respect to the jet axis.
The general form of the differential equations can be written as follows \cite{Chang:2016gjp}:
\begin{eqnarray}
\label{eq:dG/dt2}
\!\!&&\!\!\frac{d}{dt}f_i(\omega_i, k_{i\perp}^2, t)=\left(\hat{e}_i \frac{\partial}{\partial \omega_i}
 + \frac{1}{4} \hat{q}_i {\nabla_{k_\perp}^2}\right)f_i(\omega_i, k_{i\perp}^2, t)  \ \ \
\nonumber \\
\!\!&&\!\!+\sum_j\int d\omega_jdk_{j\perp}^2 \frac{d\tilde{\Gamma}_{j\rightarrow i}(\omega_i, k_{i\perp}^2|\omega_j, k_{j\perp}^2)}{d\omega_i d^2k_{i\perp}dt} f_j(\omega_j, k_{j\perp}^2, t)\nonumber \\
\!\!&&\!\!-\sum_j \int d\omega_jdk_{j\perp}^2 \frac{d\tilde{\Gamma}_{i\rightarrow j}(\omega_j, k_{j\perp}^2|\omega_i, k_{i\perp}^2)}{d\omega_j d^2k_{j\perp}dt} f_i(\omega_i, k_{i\perp}^2, t).
\end{eqnarray}
Here, the first and second terms on the right-hand side (in the first line) account for the effects of collisional energy loss and transverse momentum broadening due to elastic scatterings with the medium constituents.
The last two terms represent medium-induced radiative processes, for which we employ the splitting kernels ${d\tilde{\Gamma}_{i\rightarrow j}}/{d\omega d^2k_{\perp}dt}$ from the higher-twist jet energy loss formalism \cite{Wang:2001ifa, Majumder:2009ge}.

To solve Eq.~(\ref{eq:dG/dt2}), the vacuum shower (which we take from PYTHIA) is supplied as the initial condition.
This means we first generate the vacuum shower, then simulate the medium-induced corrections on top of the vacuum shower.
Such method has been used in many jet quenching studies, such as Refs.~\cite{Young:2011qx,Casalderrey-Solana:2014bpa,He:2015pra}.
The second widely-used method is to combine the vacuum and medium-induced splitting to a single kernel and simulate the vacuum and medium-induced radiations together simultaneously, such as Refs. \cite{Zapp:2012ak, Apolinario:2012cg,Cao:2017qpx}.
These two methods are complimentary: the second one works better for the early stage of jet evolution when jets have large
virtuality and the vacuum shower dominates, while our method (the first one) describes better the later stage
of jet evolution when jets are close to on-shell and the medium-induced radiation dominates.

In Eq.~(\ref{eq:dG/dt2}), the information of the QGP medium is encoded in the jet transport parameters: $\hat{e}$ for longitudinal momentum (energy) loss and $\hat{q}$ for transverse momentum broadening.
In this work, we relate $\hat{q}$ to the local temperature $T$ and flow four-velocity $u$ of the QGP medium as follows \cite{Chen:2010te}:
\begin{equation}
\label{eq:qhat}
\hat{q} (\tau,\vec{r}) = \hat{q}_0 \cdot \frac{T^3(\tau,\vec{}r)}{T_{0}^3(\tau_{0},\vec{0})} \cdot \frac{p\cdot u(\tau, \vec{r})}{p_0},
\end{equation}
where $T_0$ is the initial temperature at the center of the QGP medium in most central 0-10\% collisions, $p^\mu$ is the four-momentum of the propagating parton, and the factor ${p\cdot u}/{p_0}$ is to account for the flow effect in a non-static medium \cite{Baier:2006pt}.
The transport parameters $\hat{q}$ for quarks and gluons are connected by the Casimir color factors $\hat{q}_{\rm gluon}/\hat{q}_{\rm quark} = C_{\rm A}/C_{\rm F}$, and we also assume the relation, $\hat{q} = 4T \hat{e}$ \cite{Moore:2004tg,Qin:2009gw}.
Accordingly, only one transport coefficient (which we choose $\hat{q}_0$ for quarks) governs the sizes of all medium effects in Eq.~(\ref{eq:dG/dt2}), and is tuned to describe one set of jet quenching observables.
In the current study, we only include the interaction of jet with the medium in QGP phase, and the small medium effect in hadronic phase is neglected, i.e., the jet-medium interaction is turned off when the local medium temperature is lower than $T_{\rm c} = 160~{\rm MeV}$.

Equation~(\ref{eq:dG/dt2}) not only describes the evolution of the jet shower partons, but also determines the energy and momentum exchange between the shower partons with the QGP medium.
The QGP medium will respond hydrodynamically to the energy and momentum deposited by the jet shower, and collective flow can be excited along with the jet propagation.
In our coupled jet-fluid model, we describe the space-time evolution of the expanding QGP fluid together with the jet-induced flow by solving the hydrodynamic equation with source term:
\begin{eqnarray}
\label{eq:hydro}
\partial_{\mu}T_{\rm fluid}^{\mu \nu}\!\!\left(x\right)=J^{\nu}\!\!\left(x\right)\!,
\end{eqnarray}
where $T_{\rm fluid}^{\mu \nu}$ is the energy-momentum tensor of the medium fluid, and $J^{\nu}$ is the source term to describe the four dimensional energy-momentum density deposited by the jet shower.
In this study, we model the QGP as an ideal fluid in local equilibrium and assume that the energy and momentum deposited by the jet are instantaneously thermalized.
Then the source term may be constructed as follows \cite{Tachibana:2017syd}:
\begin{eqnarray}
\label{eq:source}
J^{\nu}\left(x\right)
&=&\sum_i\int
\frac{d\omega_{i}dk^{2}_{i \perp} d\phi_{i}}{2\pi}
\delta^{(3)}\left(\mathbf{x} - {\mathbf{x}}_0^{\rm jet} - \frac{\mathbf{k}_i}{\omega_i}t\right)
\nonumber\\
&\times & {k^\nu_i} \left(\hat{e}_i \frac{\partial}{\partial \omega_i}
 + \frac{1}{4} \hat{q}_i {\nabla_{k_{\perp}}^2}\right)f_i(\omega_i, k_{i\perp}^2, t)\,,
 \end{eqnarray}
where $\phi_{i}$ is the azimuth angle with respect to the jet axis.

Currently, the finite viscosities of the medium fluid are not implemented yet in our model.
Although they are essential for more precise description of the collective flows in heavy-ion collisions \cite{Song:2007fn, Schenke:2010rr, Schenke:2011tv, Song:2011qa,Petersen:2011sb, Shen:2011eg},
it is shown by semi-analytical calculations \cite{Yan:2017rku} that their effect on the angular structure of final state hadrons brought by jet-induced flow is not significant.
We would like to leave the detailed studies of the small viscous correction on the structural development of the jet-induced flow,
e.g., the smearing of the jet-induced shockwave by the finite shear viscosity, for a future work.

In our coupled jet-fluid model, Eq.~(\ref{eq:hydro}) is numerically solved in the $(3 +1)$-dimensional relativistic $\tau$-$\eta_{\rm s}$ coordinates with the source term given by Eq.~(\ref{eq:source}), which is constructed from the solutions of the jet-shower transport equations (\ref{eq:dG/dt2}).
We set up the initial condition of the medium fluid at $\tau=0.6~{\rm fm/}c$ by applying the optical Glauber$+$the modified BGK model \cite{Hirano:2005xf}.
The parameters in the initial condition model for Pb+Pb collisions at 2.76~ATeV and 5.02~ATeV are chosen to reproduce the pseudorapidity density distribution for charged particles measured by the ALICE Collaboration \cite{Adam:2015kda, Adam:2016ddh}.
Here, we do not consider the geometrical fluctuation of the nucleons and their internal structures in the incident heavy ions to reduce computational cost.
Although the event-by-event fluctuations of the initial conditions have significant effect on our understanding of many flow observables, they have moderate effect to jet suppression (see e.g., Ref. \cite{Noronha-Hostler:2016eow}). The inclusion of the initial state fluctuations could change a little our extracted value of $\hat{q}$, but should not affect the conclusion drawn in this work.
For the equation of state of the medium fluid, we employ the parameterizations of the lattice QCD calculation in Ref.~\cite{Borsanyi:2013bia}.
Along with the hydrodynamic expansion, the medium fluid cools down and eventually turns from QGP to hadronic matter according to the equation of state.
The evolution of the hadronic matter fluid stops when the freeze-out is accomplished.
In this study, we assume the isothermal freeze-out and set the freeze-out temperature $T_{\rm FO}=140~{\rm MeV}$.

After finishing the in-medium evolution of the partons in jet shower according to Eq.~(\ref{eq:dG/dt2}), we obtain their contribution to the transverse momentum of the reconstructed jet with a given cone size $R$ as follows:
\begin{equation}
\label{eq:eshower}
p_{\rm T}^{\rm shower}(R) = \sum_i \int_R
d\omega_i dk_{i\perp}^2
\omega_i f_i(\omega_i, k_{i\perp}^2),
\end{equation}
where the subscript $R$ means that the integral is taken with the constraint, $k_{i,\perp}/\omega_i<R$.
In our coupled jet-fluid model, hadron spectra from the medium with the hydrodynamic medium response are obtained via the Cooper-Frye formula \cite{Cooper:1974mv} as in the conventional hydrodynamic models (further details are found in Ref.~\cite{Tachibana:2017syd}).
The pure contribution from hydrodynamic medium response is estimated by subtracting out the hadron spectra of the background medium (without jet propagation):
\begin{eqnarray}
\frac{d \Delta N^{\rm hydro}}{d^3 p} &=&
\left.\frac{dN^{\rm hydro}}{d^3 p}\right|_{\mbox{\footnotesize w\!/ \!jet}} - \left. \frac{dN^{\rm hydro}}{d^3 p}\right|_{\mbox{\footnotesize w\!/\!o \!jet}}.
\label{eq:subtract}
\end{eqnarray}
To construct the jet including hydrodynamic medium response effect, the contribution of ${d\Delta N^{\rm hydro}/d^3 p}$ is added to the parton shower part in Eq.~(\ref{eq:eshower}):
\begin{eqnarray}
\label{eq:ejet}
p_{\rm T}^{\rm jet}(R) &=& p_{\rm T}^{\rm shower}(R) + p_{\rm T}^{\rm hydro}(R), \\
\label{eq:ehyd}
p_{\rm T}^{\rm hydro}(R) &=&
\int_R d^3p p_{\rm T} \frac{d\Delta N^{\rm hydro}}{d^3 p},
\end{eqnarray}
where the integral for the medium response part is taken for the region $\sqrt{(\eta_{\rm p}-\eta_{\rm p}^{\rm jet})^2+(\phi_p-\phi_p^{\rm jet})^2}<R$.

\section{Results and analysis}
\label{results}

\begin{figure}[tbhp]
\begin{center}
\centering
\includegraphics[width=1.0\linewidth]{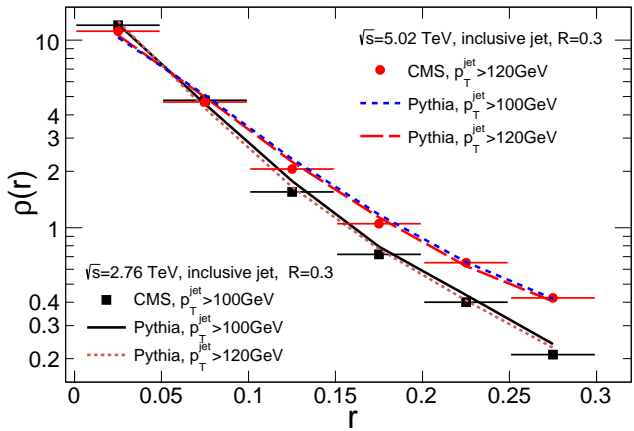}
\caption{Jet shape function in $p$+$p$ collisions for jets at 2.76~ATeV and 5.02~ATeV, compared with data from the CMS Collaboration \cite{Chatrchyan:2013kwa,Sirunyan:2018jqr}.}
\label{fig:initial}
\end{center}
\end{figure}

In this section, we present the results of full jet observables from the numerical simulations of Pb+Pb collisions at 2.76~ATeV and 5.02~ATeV using our coupled jet-fluid model.
In the simulations, the jet production points in the transverse plane $\eta_{\rm s} = 0$ are generated according to the distributions of the binary nucleon-nucleon collisions calculated by using a Glauber model simulation \cite{Miller:2007ri, Qin:2010pf}.
The initial jet spectrum is obtained via PYTHIA simulation \cite{Sjostrand:2007gs}, and the FASTJET \cite{Cacciari:2011ma} package is employed for full jet reconstruction in PYTHIA.
The jets are assumed to be created at $t=0$ and free-stream without interaction until the thermalization proper time of the QGP, $\tau=0.6~{\rm fm/}c$.
Then, the jet shower and the QGP fluid interact with each other and evolve according to the jet shower transport equations~(\ref{eq:dG/dt2}) and the hydrodynamic equations with source terms~(\ref{eq:hydro}).

To solve the coupled differential transport equations~(\ref{eq:dG/dt2}) for jet shower evolution, the initial conditions for quark and gluon three-dimensional momentum distribution have to be provided.
We generate them using PYTHIA \cite{Sjostrand:2006za} with a parameter set tuned to reproduce the jet shape function in $p$+$p$ collisions.
The jet shape function is the radial direction distribution of the transverse momentum inside jets and defined as follows:
\begin{eqnarray}
\label{eq:rho_r}
\rho_{\rm jet}(r) =\frac{1}{\delta r} \frac{\sum_{{|r_i - r|\leq\frac{1}{2}\delta r}} p_{\rm T}^i}{\sum_{r_i<R} p_{\rm T}^i} ,
\end{eqnarray}
where $r_i = \sqrt{(\eta^i_{\rm p} - \eta^{\rm jet}_{\rm p})^2 + (\phi_p^i - \phi_p^{\rm jet})^2}$, $\delta r$ is the bin size, and the sum over $i$ runs over constituents of the full jets.
Figure~\ref{fig:initial} shows the baseline of the jet shape function obtained from the PYTHIA simulations, compared to the experimental data in $p$+$p$ collisions measured by the CMS Collaboration \cite{Chatrchyan:2013kwa,Sirunyan:2018jqr}.
One can see that the jet shape function for jets with $p_{\rm T}>100$~GeV at 2.76~TeV $p$+$p$ collisions is steeper than that at 5.02~TeV.
Such collision energy dependence of the initial jet shape can lead to some difference between the medium modifications of jet shapes in Pb+Pb collisions at 2.76~ATeV and 5.02~ATeV, which is illustrated in a later subsection.

\subsection{Jet $R_{AA}$}
\label{eloss}

We first fix the parameter $\hat{q}_0$ in our model by comparison with the measurements of nuclear modification factor $R_{\rm AA}$ of single inclusive jet spectra, defined as:
\begin{eqnarray}
R_{\rm AA} &=&
\frac{1}{\langle N_{\rm coll}\rangle}
\frac
{d^2 N^{\rm AA}_{\rm jet} / d\eta^{\rm jet}_{\rm p} d p^{\rm jet}_{\rm T}}
{d^2 N^{pp}_{\rm jet} / d\eta^{\rm jet}_{\rm p} d p^{\rm jet}_{\rm T}},
\end{eqnarray}
where $\langle N_{\rm coll}\rangle$ is the average number of binary nucleon-nucleon collisions in a given centrality class, $N^{\rm AA}_{\rm jet}$ is the number of jets in nucleus-nucleus collisions, and $N^{pp}_{\rm jet}$ is that in $p$+$p$ collisions.

\begin{figure}
\begin{center}
\centering
\includegraphics[width=1.0\linewidth]{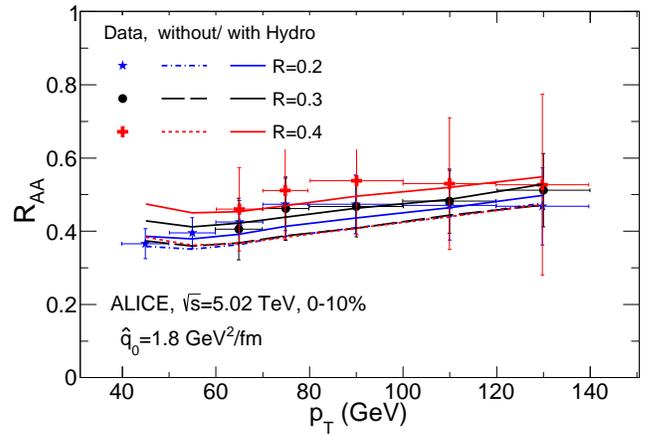}
\caption{Nuclear modification factor $R_{\rm AA}$ of single inclusive jet production with and without taking into account the contribution from hydrodynamic medium response in central Pb+Pb collisions at 5.02A TeV for jet cone sizes $R=0.2,0.3,$ and $0.4$.
Experimental data are taken from the ALICE Collaboration \cite{Hosokawa:2019odr}.
}
\label{fig:raa-R}
\end{center}
\end{figure}

Figure~\ref{fig:raa-R} shows our results for single inclusive jet $R_{\rm AA}$ in Pb+Pb collisions at 5.02~ATeV with different jet cone sizes, compared with the experimental data from the ALICE Collaboration \cite{Hosokawa:2019odr}.
Our full results with hydrodynamic medium response effect agree reasonably with the data within the experimental errors.
Here, we set $\hat{q}_0 = 1.8~{\rm GeV^2/fm}$ for quarks, and use this value for Pb+Pb collisions at 5.02~ATeV throughout this paper.
As a general feature, one can see that the inclusion of the hydrodynamic medium response effect rises the value of $R_{\rm AA}$, which is simply because some part of energy lost by the parton shower is now recovered.
Also, a clear jet cone size dependence is observed after the inclusion of the hydrodynamic medium response effect.
A particularly interesting feature is that while most of the energy in the shower part is well collimated and can be captured by a narrow jet cone, the energy carried by the medium response effect spreads widely around the jet axis \cite{Tachibana:2017syd}.
Thus, sizable energy can be gained from the medium response contribution when the jet cone size is increased.
The same trend of the jet cone size dependence can also be seen in the ALICE data although the experimental error bars are still too large to draw a firm conclusion.

\begin{figure}
\begin{center}
\centering
\includegraphics[width=1.0\linewidth]{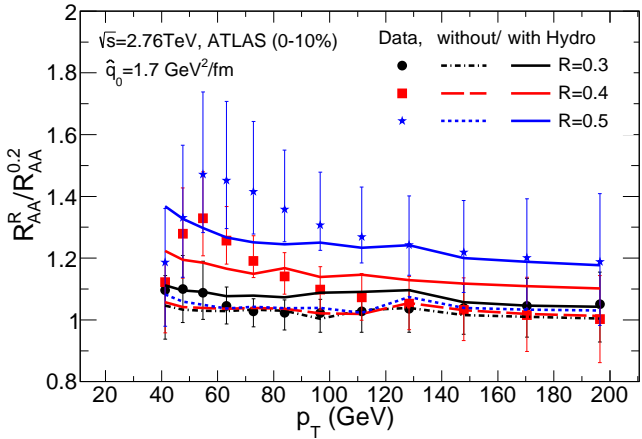}
\caption{Ratio of single inclusive jet $R_{\rm AA}$ with jet cone sizes $R=0.3$-$0.5$ to that with jet cone size $R=0.2$ in central Pb+Pb collisions at 2.76~ATeV.
	Experimental data are from the ATLAS Collaboration \cite{Aad:2012vca}.}
\label{fig:raa-R2}
\end{center}
\end{figure}

Clearer jet cone size dependence can be seen in the ATLAS data in Fig.~\ref{fig:raa-R2} which shows the ratios of the single inclusive $R_{\rm AA}$ for jets with cone sizes $R=0.3$-$0.5$ to that with $R=0.2$ in Pb+Pb collisions at 2.76~ATeV.
For the simulations of jet events in Pb+Pb collisions at 2.76~ATeV, we set $\hat{q}_0 = 1.7~{\rm GeV^2/fm}$ for quarks, which is obtained to reproduce the data for single inclusive jet $R_{\rm AA}$ taken from the CMS Collaboration \cite{Khachatryan:2016jfl} in our previous work \cite{Tachibana:2017syd}.
As shown by our calculation as well as the ATLAS data, when the contribution of the hydrodynamic medium response is taken into account, jet energy loss decreases when increasing the jet cone size.
Such jet cone size dependence of jet energy loss and jet quenching has also been found in previous studies \cite{Vitev:2009rd, Zapp:2012ak, Tachibana:2017syd,He:2018xjv}.

\subsection{$\gamma$-jet asymmetry}
\label{XJ}

Jets tagged with isolated photons, known as $\gamma$-jets, have attracted a lot of interests in jet quenching studies \cite{Wang:1996yh,Zhang:2009rn,Qin:2012gp,Dai:2012am,Chatrchyan:2012gt,Wang:2013cia,Luo:2018pto}.
Since the triggered photons do not interact with the QGP medium after they are produced, one just needs to focus on the medium effect on the away-side jets in relativistic heavy-ion collisions.
Due to the next-to-leading order effect, the away-side jets typically have different transverse momenta from the triggered photons when they are first produced in vacuum (p+p collisions).
Such transverse momentum imbalance is usually quantified by using the momentum fraction variable $X_{\rm J\gamma}\equiv p_{\rm T}^{\rm jet}/p_{\rm T}^{\gamma}$.
The medium effect due to the jet energy loss in heavy-ion collisions will manifest as the modification to the event distribution of $X_{\rm J\gamma}$, i.e., $({1}/{N_{\rm J \gamma}}) {dN_{\rm J \gamma}}/{dX_{\rm J\gamma}}$.

\begin{figure}[h]
\begin{center}
\centering
\includegraphics[width=\linewidth]{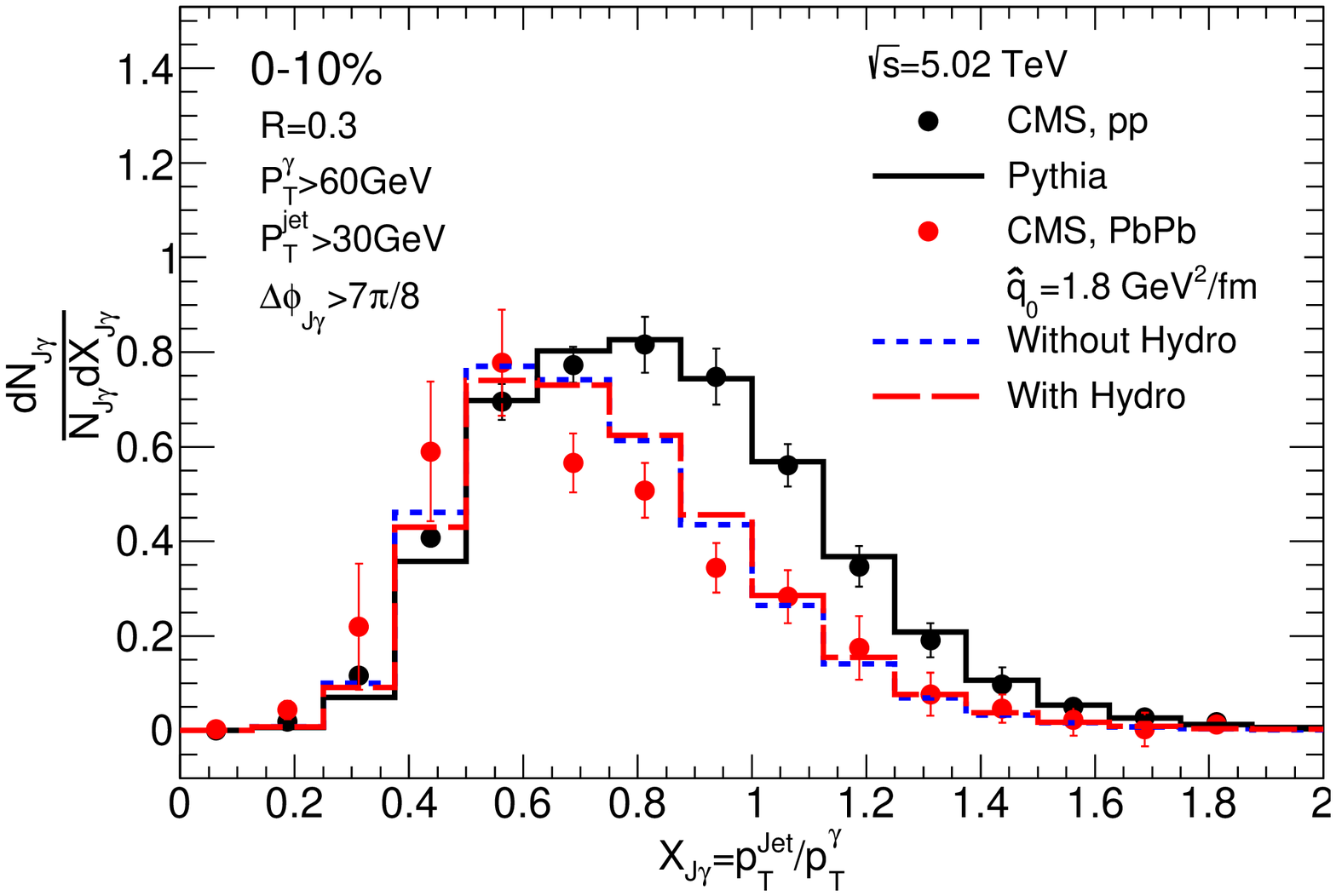}
\includegraphics[width=\linewidth]{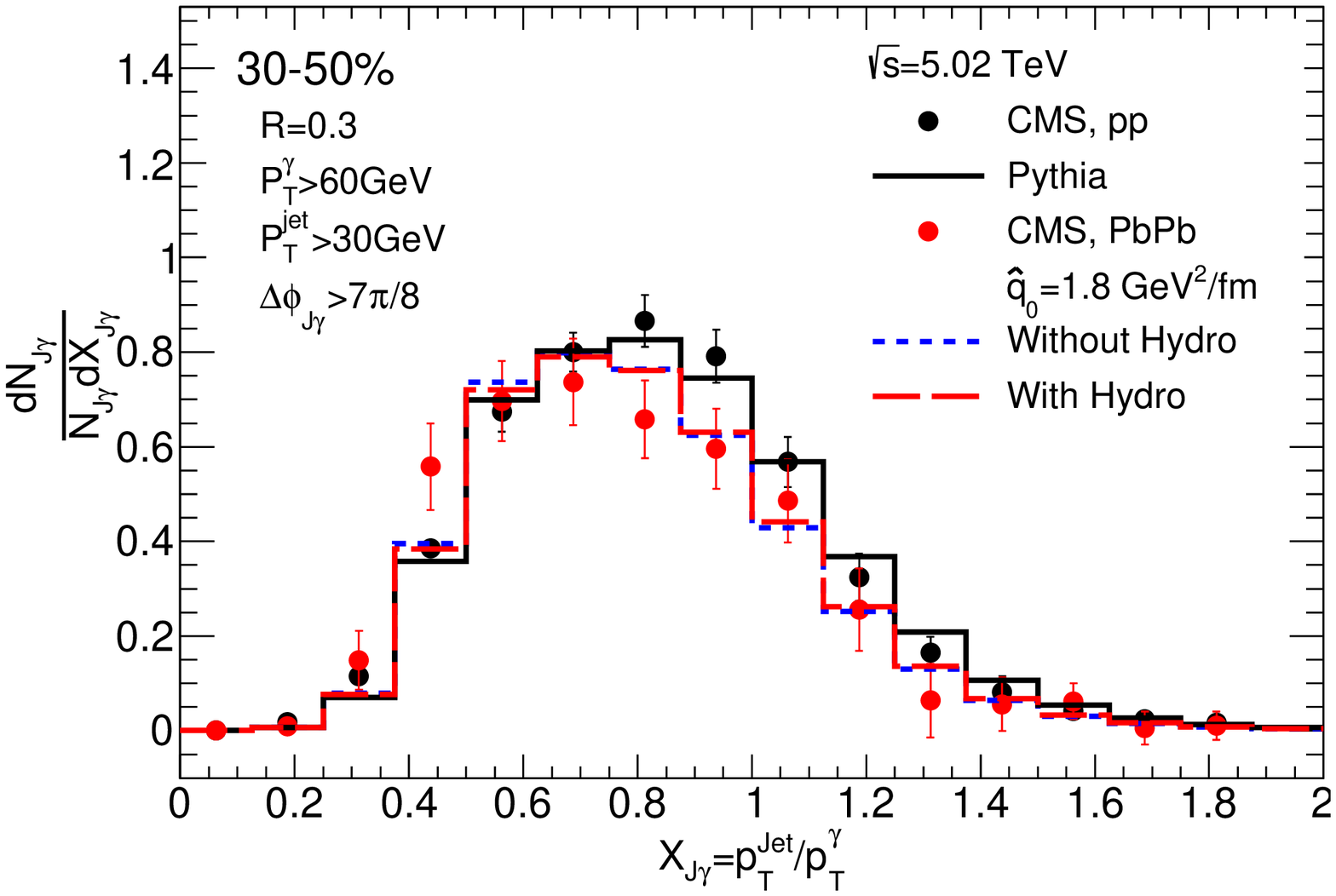}
\caption{Event distribution of $\gamma$-jet momentum fraction $X_{\rm J\gamma}$ in $p$+$p$ collisions and in $0$-$10\%$ centrality (upper) and $30$-$50\%$ centrality (lower) Pb+Pb collisions at 5.02~ATeV.
The trigger threshold of the transverse momentum is set to $p^{\gamma}_{\rm T}>60~{\rm GeV}$ for photons and $p^{\rm jet}_{\rm T}>30~{\rm GeV}$ for jets.
The relative azimuthal angle between photon and jet $\Delta \phi_{\rm J\gamma}$  is required to be larger than $7\pi/8$.
Results with and without the contribution of hydrodynamic medium response are shown for Pb+Pb collisions.
Experimental data are taken from the CMS Collaboration \cite{Sirunyan:2017qhf}.
}
\label{fig:gamma}
\end{center}
\end{figure}

Figure~\ref{fig:gamma} shows the nuclear modification of $X_{\rm J\gamma}$ distribution in Pb+Pb collisions at 5.02~ATeV.
Our model provides a reasonable description of the $X_{\rm J\gamma}$ distribution and its shift towards smaller $X_{\rm J\gamma}$ in Pb+Pb collisions due to the jet energy loss.
It is noteworthy that the effect of hydrodynamic medium response in the $X_{\rm J\gamma}$ distribution is not as obvious as in the single inclusive jet $R_{\rm AA}$.

\subsection{Jet shape}
\label{shape}

The single inclusive jet $R_{\rm AA}$ and the nuclear modification of $X_{\rm J\gamma}$ distribution provide the information on the overall amount of jet energy loss.
More detailed information about the medium modification on jet shower evolution and the medium response to jet transport can be obtained by studying the jet structure observables which are more sensitive to the details on the redistribution of the energy among the constituents of the jet after traversing the QGP medium.
Jet shape function defined in Eq.~(\ref{eq:rho_r}) is one of common jet structure observables and describes the transverse energy profile of jets.

\begin{figure}
\begin{center}
\centering
\includegraphics[width=\linewidth]{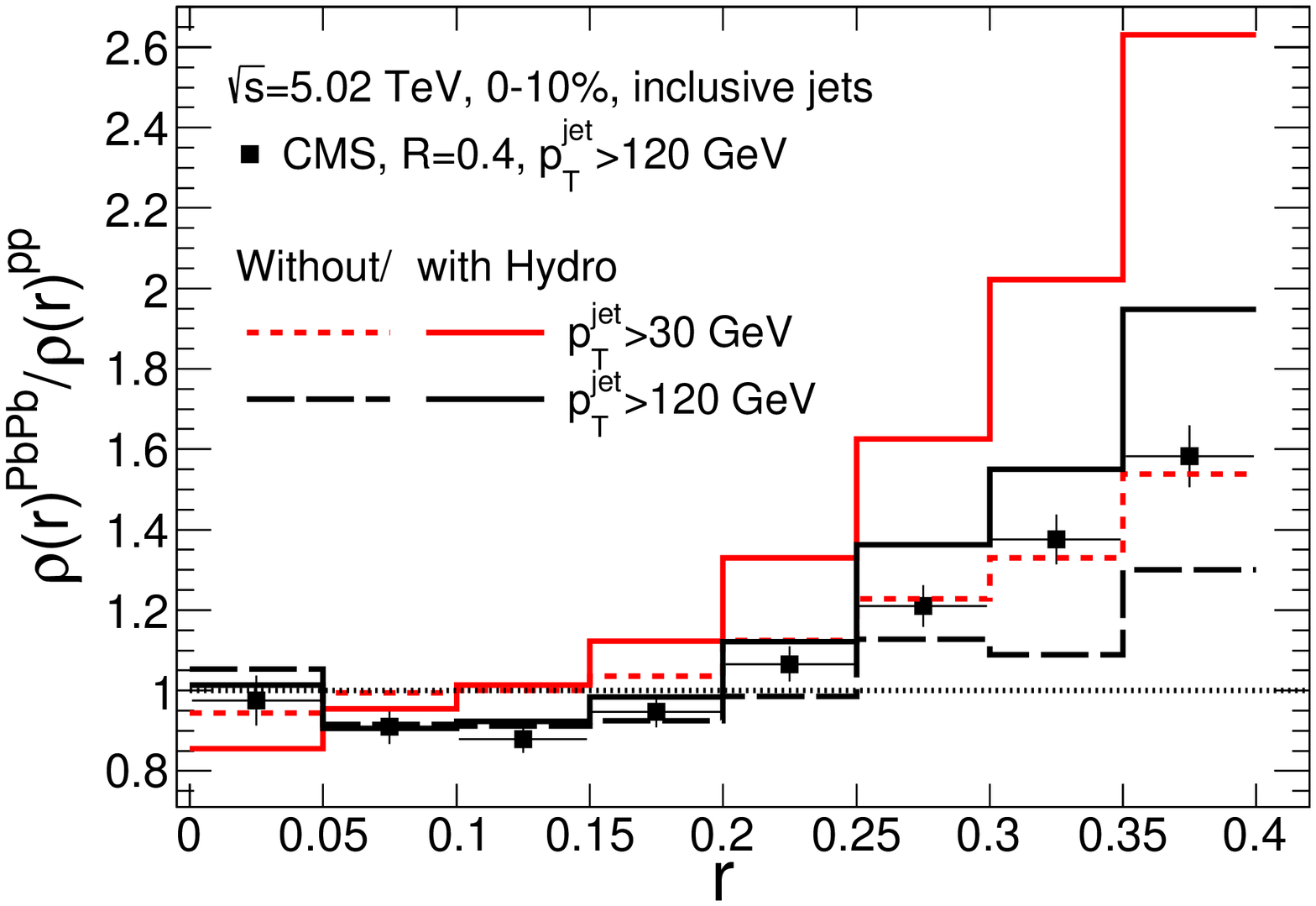}
\includegraphics[width=\linewidth]{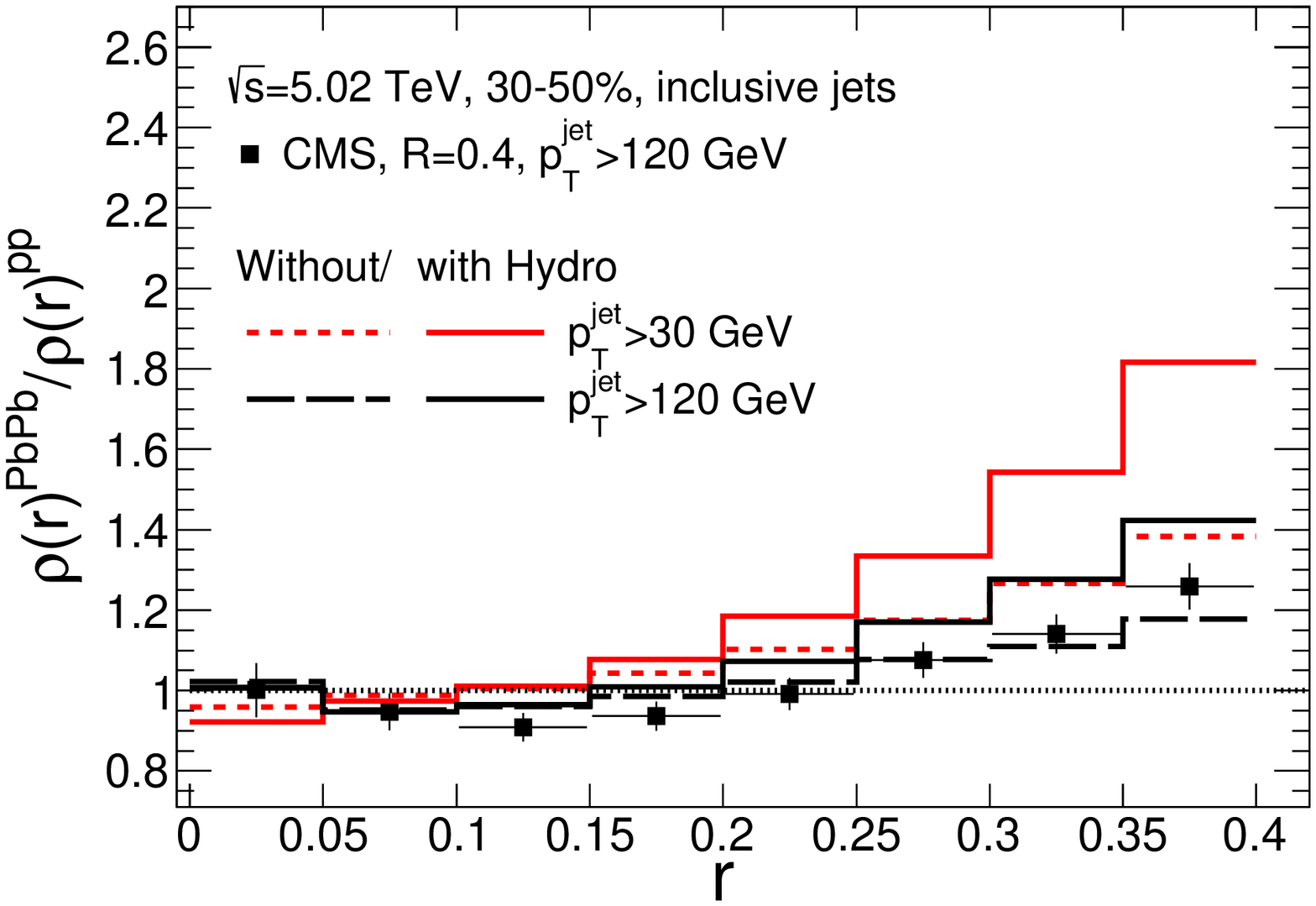}
\caption{Nuclear modification factor of jet shape function for single inclusive jets with cone size $R=0.4$ in $0$-$10\%$ centrality (upper) and  $30$-$50\%$ centrality (lower) Pb+Pb collisions at 5.02~ATeV.
Results with and without the contribution of hydrodynamic medium response are shown for jet transverse momentum cuts $p^{\rm jet}_{\rm T}>30~{\rm GeV}$ and $p^{\rm jet}_{\rm T}>120~{\rm GeV}$.
Experimental data for jets with $p^{\rm jet}_{\rm T}>120~{\rm GeV}$ are taken from the CMS Collaboration \cite{Sirunyan:2018jqr}.}
\label{fig:rho}
\end{center}
\end{figure}

Figure~\ref{fig:rho} shows the nuclear modification factor of jet shape function for inclusive jets in Pb+Pb collisions at 5.02~ATeV.
One can see that our results for $p^{\rm jet}_{\rm T}>120~{\rm GeV}$ both with and without the hydrodynamic response effect can capture the typical features in the modification pattern measured by the CMS Collaboration:
an enhancement at large $r$ and a suppression of the ratio at small $r$.
It is interesting that the suppression (dip) here is at $r \sim 0.1$, not at $r\sim 0$, i.e., the ratio is a non-monotonic function of $r$.
Such non-monotonic dependence on $r$ indicates a collimation of jet energy toward the jet axis (the inner core part at small $r$) together with the broadening of the outer tail part (at large $r$).
This modification pattern can be naturally explained by a combination of various jet-medium interaction mechanisms \cite{Chang:2016gjp}.
While both transverse momentum broadening and medium-induced radiation transfer energy from the center to the periphery of the jet, collisional energy loss (absorption) makes the jet narrower since the soft partons in the jet periphery are more easily absorbed by the medium.
The contribution from hydrodynamic medium response does not modify very much the jet shape at small $r$, but gives rise to a significant additional enhancement of the broadening at large $r$.
All these features in jet shape modification are weaker in more peripheral collisions due to smaller size and lower temperature of the QGP medium.
Our predictions on the jet shape modification for single inclusive jets with lower transverse momentum cut $p_{\rm T}^{\rm jet}>30~{\rm GeV}$ show monotonic broadening behavior, which is drastically different from that for $p_{\rm T}^{\rm jet}>120~{\rm GeV}$.
This transverse momentum dependence in jet shape modification has been predicted by our previous work~\cite{Chang:2016gjp}.
Also, the effect of hydrodynamic medium response presents stronger enhancement at large $r$ for jets with lower jet transverse momentum.

\begin{figure}
\begin{center}
\centering
\includegraphics[width=\linewidth]{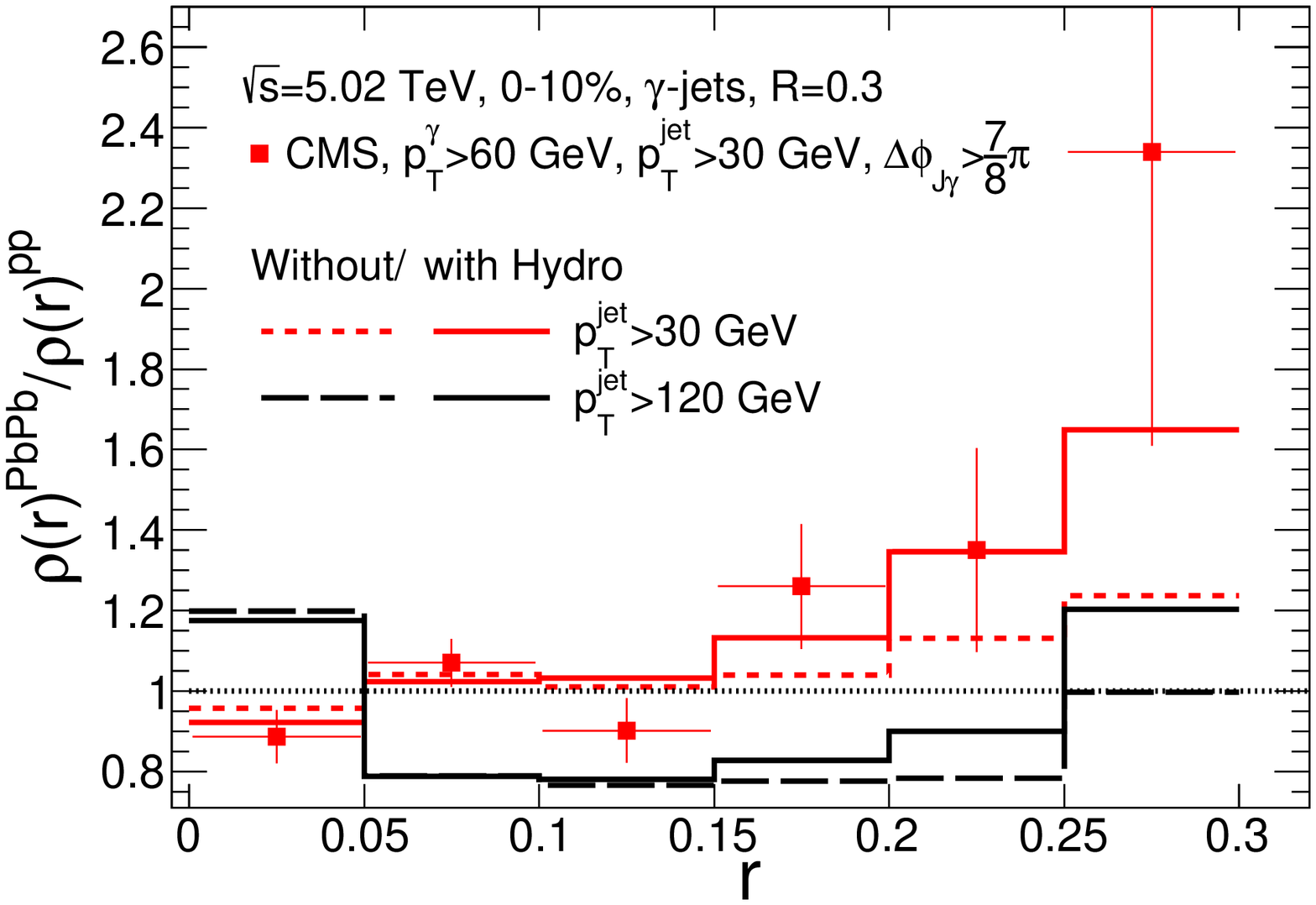}
\includegraphics[width=\linewidth]{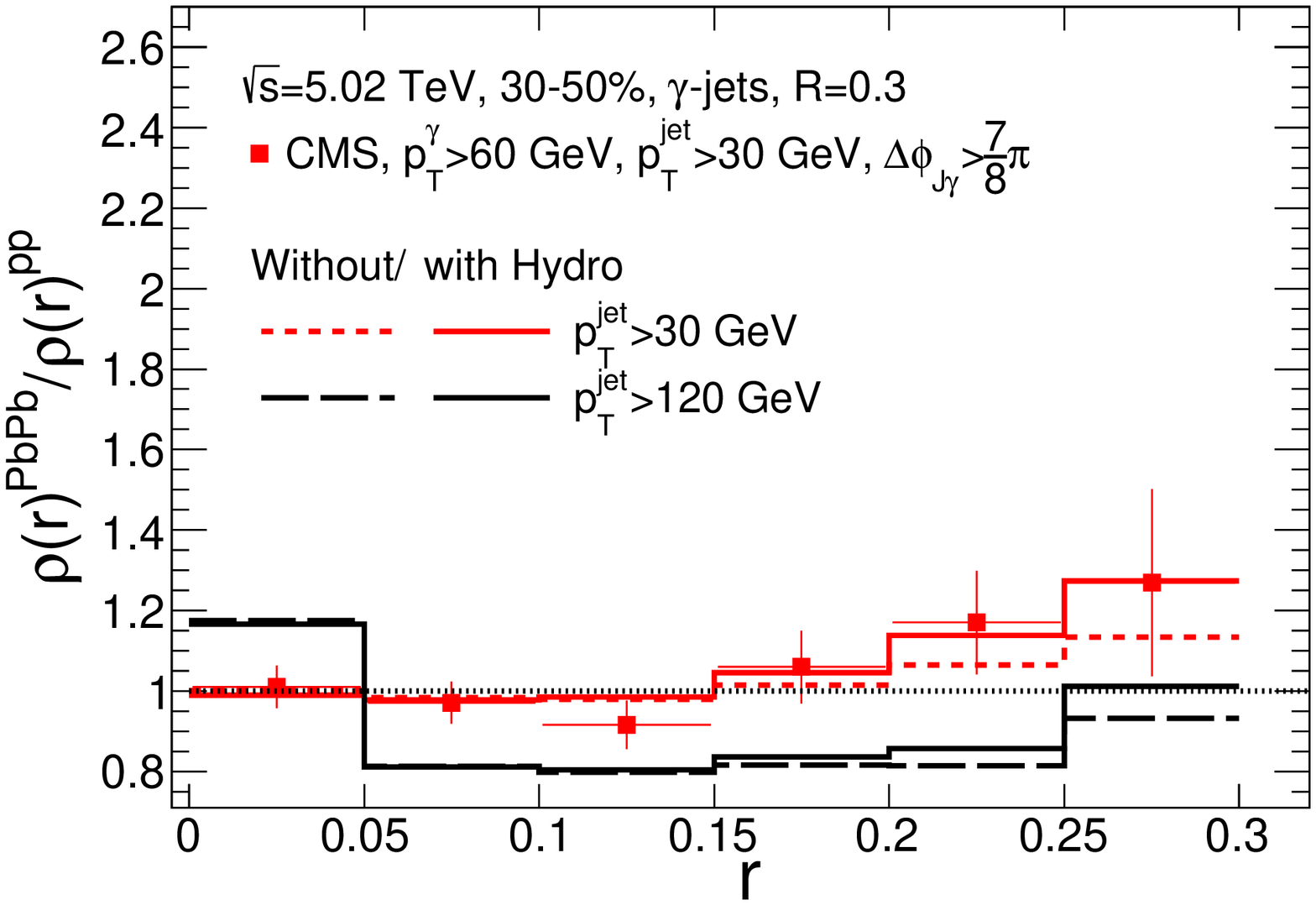}
\caption{Nuclear modification factor of jet shape function for $\gamma$-jets with cone size $R=0.3$ in $0$-$10\%$ centrality (upper) and $30$-$50\%$ centrality (lower) Pb+Pb collisions at 5.02~ATeV.
	The photon has transverse momentum $p^{\gamma}_{\rm T}>60~{\rm GeV}$, and the relative azimuthal angle between photon and jet is $\Delta \phi_{\rm J\gamma}>7\pi/8$.
	Results with and without the contribution of hydrodynamic medium response are shown for jet transverse momentum cuts $p^{\rm jet}_{\rm T}>30~{\rm GeV}$ and $p^{\rm jet}_{\rm T}>120~{\rm GeV}$.
	Experimental data for jets with $p^{\rm jet}_{\rm T}>30~{\rm GeV}$ are taken from the CMS Collaboration \cite{Sirunyan:2018ncy}.
}
  \label{fig:rho-gamma}
\end{center}
\end{figure}

Figure~\ref{fig:rho-gamma} shows the nuclear modification factor of jet shape function for $\gamma$-jets in Pb+Pb collisions at 5.02~ATeV.
Our results for $p^{\rm jet}_{\rm T}>30~{\rm GeV}$ show the monotonic increase as a function of $r$ due to the broadening effect, which agrees reasonably well with the experimental data \cite{Sirunyan:2018ncy}.
This pattern (i.e., the monotonic increase as a function of $r$) is different from that of inclusive jets with $p^{\rm jet}_{\rm T}>120~{\rm GeV}$ which shows a clear dip at $r\sim 0.1-0.15$ \cite{Sirunyan:2018jqr}.
It has been argued that such difference is due to the different parton flavor compositions (quark or gluons) in $\gamma$-jets and single inclusive jets \cite{Chien:2015hda}.
However, our results for $\gamma$-jets with $p^{\rm jet}_{\rm T}>120~{\rm GeV}$ also have a clear dip structure, which is similar to single inclusive jets with $p^{\rm jet}_{\rm T}>120~{\rm GeV}$.
This indicates that whether the dip structure appears or not is more determined by jet transverse momenta.

As is known, jet shape function is a deep falling function of $r$.
This means that a large fraction of energy of the jet is contained in a very small range of $r$ and the outer part of the jet contains just a small fraction of the jet energy.
Also, Fig.\ref{fig:initial} shows that jet shape function for lower energy jets is less deep falling, which means that less fraction of jet energy is contained in the inner part of the jet.
Our study shows that for lower energy jets, the broadening effects via medium-induced radiation and transverse momentum broadening are stronger \cite{Chang:2016gjp}, which leads to the suppression of the jet shape function at around $r=0$.
Therefore, the nuclear modification of jet shape function (for smaller jet energies) increases monotonically as a function of $r$.

\begin{figure}
\begin{center}
\centering
\includegraphics[width=\linewidth]{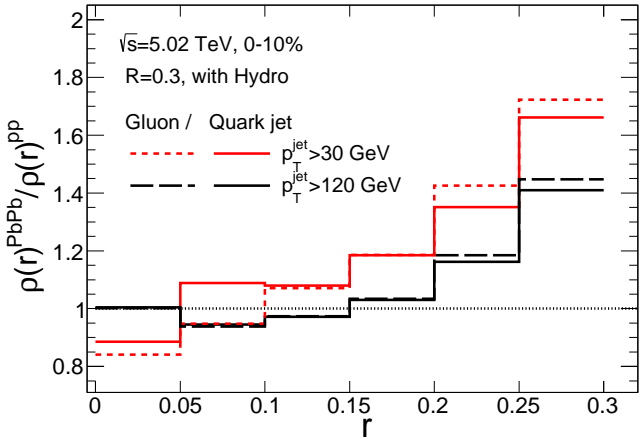}
\caption{Nuclear modification factor of jet shape function for inclusive quark jets and inclusive gluon jets with cone size $R=0.3$ in central Pb+Pb collisions at 5.02~ATeV.
Results with and without the contribution of hydrodynamic medium response are shown for jet transverse momentum cuts $p^{\rm jet}_{\rm T}>30~{\rm GeV}$ and $p^{\rm jet}_{\rm T}>120~{\rm GeV}$.
}
\label{fig:flavor}
\end{center}
\end{figure}

To illustrate that the modification pattern of jet shape mainly depends on the jet transverse momentum rather than the jet flavor, we show in Fig.~\ref{fig:flavor} the nuclear modification factors of jet shape functions separately for quark jets and gluon jets.
One can see that the jet shape modification patterns are almost the same for quark jets and gluon jets.
For large transverse momentum jets, one can see both the collimation of the jet at small $r$ and the broadening of the jet at large $r$.
For low transverse momentum jets, the monotonic broadening behavior is observed.
These features are similar to both inclusive jets and $\gamma$-jets.

\begin{figure}
\begin{center}
\centering
\includegraphics[width=\linewidth]{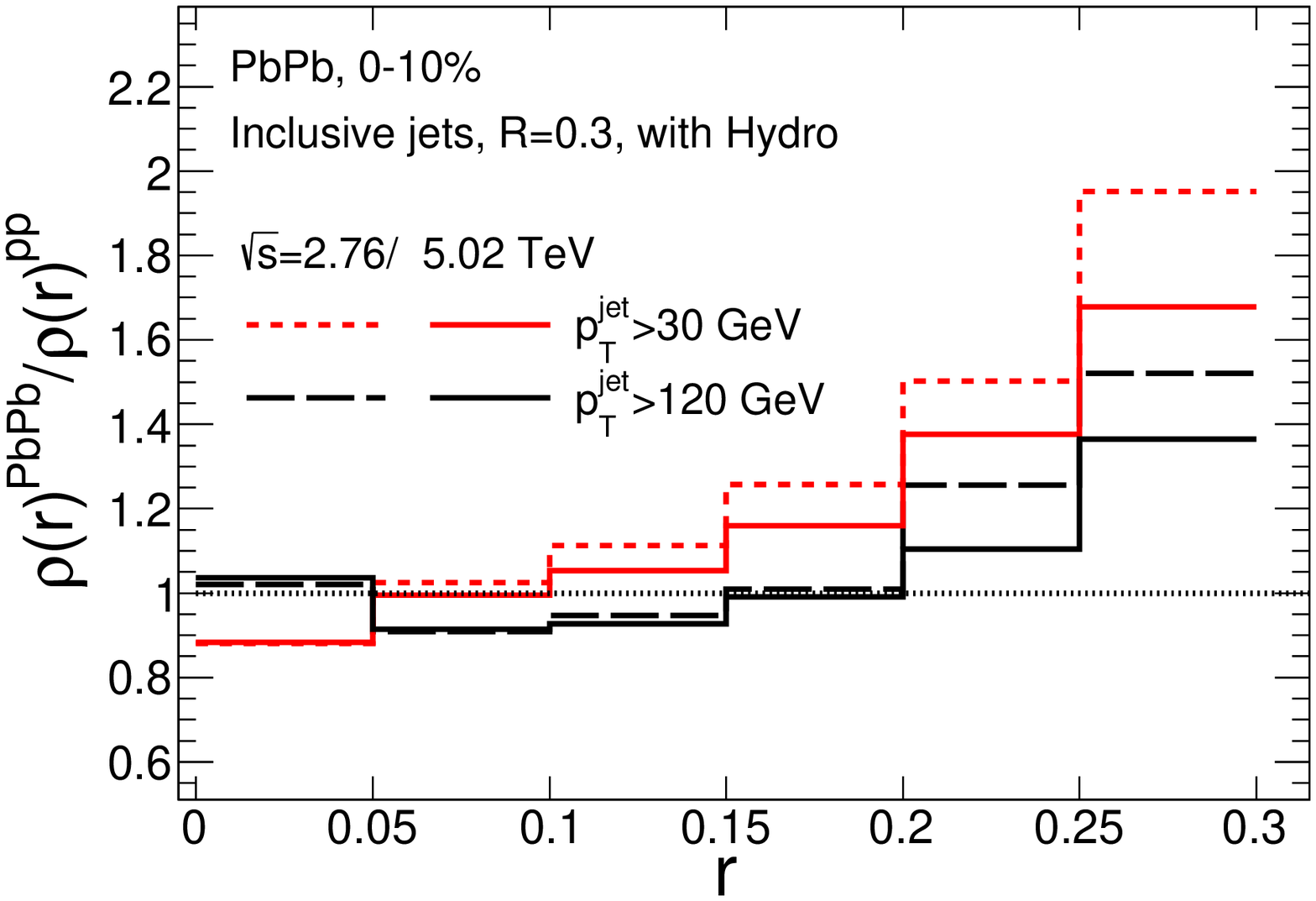}
\includegraphics[width=\linewidth]{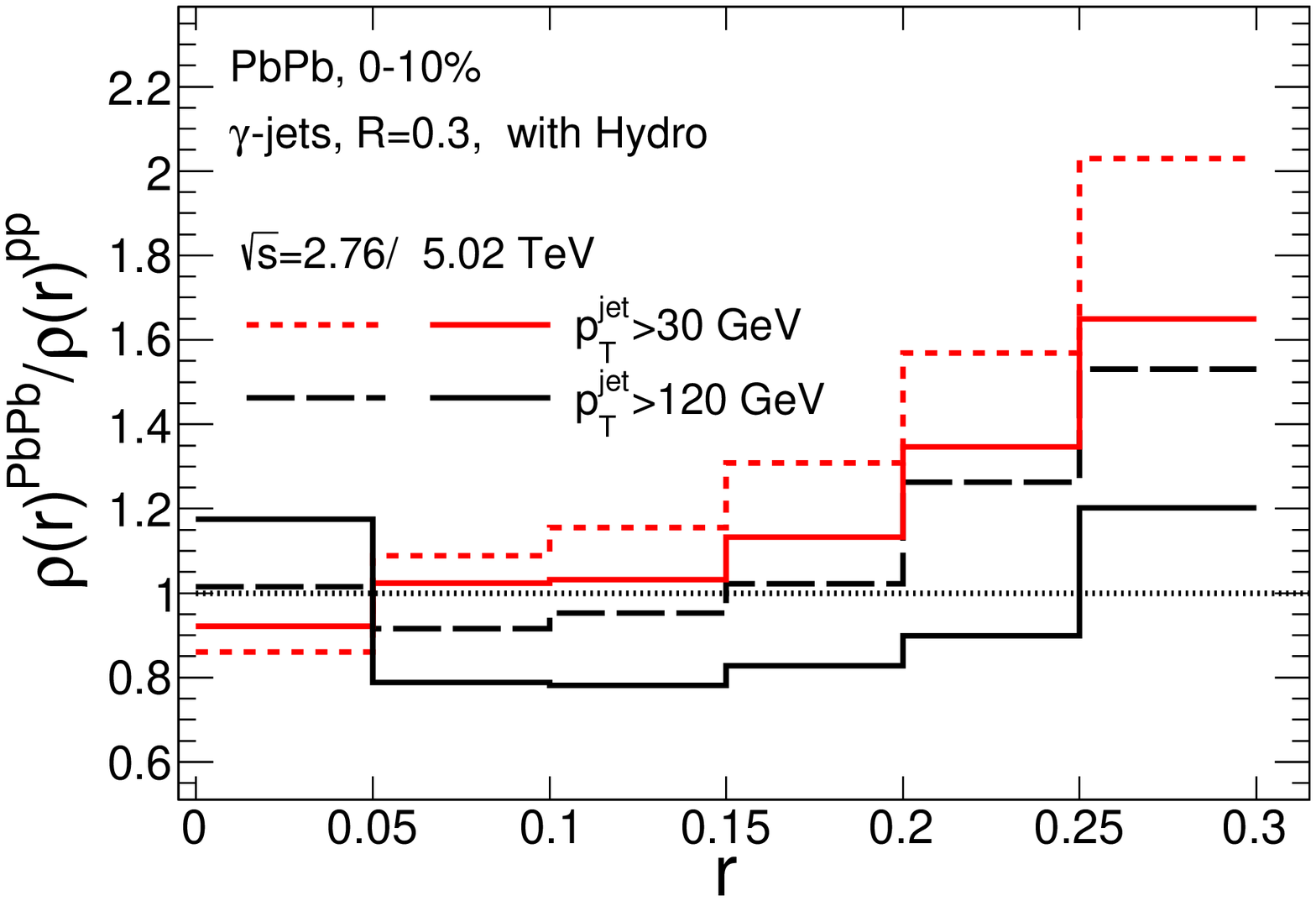}
\caption{Nuclear modification factor of jet shape function for inclusive jets (upper) and $\gamma$-jets (lower) with cone size $R=0.3$ in central Pb+Pb collisions at 2.76~ATeV and 5.02~ATeV.
Results with and without the contribution of hydrodynamic medium response are shown for jet transverse momentum cuts $p^{\rm jet}_{\rm T}>30~{\rm GeV}$ and $p^{\rm jet}_{\rm T}>120~{\rm GeV}$.
}
  \label{fig:s}
\end{center}
\end{figure}

We finally present the collision energy dependence of the jet shape modification for central Pb+Pb collisions in Fig.~\ref{fig:s}.
While the basic trends of the jet shape modification are the same at both 2.76~ATeV and 5.02~ATeV, there is sizable difference in large $r$ region.
Since jets produced at 5.02~ATeV are broader than those at 2.76~ATeV (see Fig.~\ref{fig:initial}), we find smaller broadening effect at larger $r$ for Pb+Pb collisions at 5.02~ATeV.

\section{Summary}
\label{sec:summary}

In this paper, we have studied the full jet modifications in Pb+Pb collisions at 2.76 ATeV and 5.02 ATeV using our coupled jet-fluid model.
The evolution of parton shower in the QGP medium is described by a set of coupled differential transport equations for the three-dimensional momentum distributions of shower partons in the jet.
The collisional energy loss, transverse momentum broadening, and medium-induced partonic radiation are taken into account for both leading and radiated partons.
The space-time evolution of the energy and momentum deposited by parton shower to the medium, together with the evolution of the bulk medium, is described  by the relativistic ideal hydrodynamic equation with source term.
The source term represents the energy-momentum transferred from the parton shower to the medium, and is constructed from the solutions of the parton shower transport equations.
The final jets in our model include the contributions from both hard jet shower part and soft medium response effect, the later is obtained via the Cooper-Frye formula after subtracting the background (without jet).

Based on the simulations with the coupled jet-fluid model, we have calculated various full jet observables for single inclusive jets and $\gamma$-jets.
For the single inclusive jet $R_{AA}$, we see the sizable contribution from the hydrodynamic medium response, which partially compensates the energy loss of the jet shower.
In particular, we find remarkable jet cone size dependence due to the hydrodynamic medium response.
The jet cone size dependence in jet $R_{AA}$ can indeed be seen in experimental measurements in Pb+Pb collisions at 5.02~ATeV by ALICE and at 2.76~ATeV by ATLAS, which can be well described by our full model calculations.
Our calculation for the $\gamma$-jet asymmetry distribution at 5.02~ATeV also agrees with the CMS data, and we find that the medium response effect gives small contribution to $\gamma$-jet asymmetry.

We have also presented a systematic study on the nuclear modification of jet shape function for different jet flavors, jet transverse momentum cuts and collision energies.
In all cases, the hydrodynamic response effect contributes additional jet broadening in the large-$r$ region ($r>0.15$-$0.25$).
Our study shows that the overall pattern of the jet shape modification is sensitive to jet transverse momenta.
For large jet transverse momentum ($p^{\rm jet}_{\rm T}>120$), the nuclear modification factor of jet shape has a clear dip structure due to the collimation around the inner hard core of jet and the enhancement at larger $r$ due to the broadening effect.
However, for small jet transverse momentum ($p^{\rm jet}_{\rm T}>30$~GeV), we observe the monotonic broadening behavior for inclusive jets, $\gamma$-jets, quark-initiated jets and gluon-initiated jets.
Our model calculations can reasonably describe the experimental data for both inclusive jets with $p^{\rm jet}_{\rm T}>120$~GeV \cite{Sirunyan:2018jqr} and $\gamma$-jets with $p^{\rm jet}_{\rm T}>30$~GeV.
This indicates that the different nuclear modification patterns for jet shape functions of single inclusive jets and $\gamma$-jets seen by CMS Collaboration may be naturally explained by different jet energies in these two measurements.
Our predictions may be tested by further experimental analysis with different jet transverse momentum cuts, in particular, using lower $p_{\rm T}^{\rm jet}$ cut for inclusive jets and higher $p_{\rm T}^{\rm jet}$ cut for $\gamma$-jets.

\section*{Acknowledgments}

This work is supported in part by the Natural Science Foundation of China (NSFC) under Grant Nos.~1775095, 11890711, 11375072 and 11905180.
N-B. C. is supported by Nanhu Scholar Program for Young Scholars of XYNU and the CCNU-QLPL Innovation Fund (Grant No. QLPL2016P01).
The work by Y.T. is supported in part by a special award from the Office of the Vice President of Research at Wayne State University and in part by the National Science Foundation (NSF) within the framework of the JETSCAPE collaboration under Award No. ACI-1550300.

\bibliographystyle{h-physrev5}

\begin{thebibliography}{10}
\bibitem{Wang:1991xy}
X.-N. Wang and M.~Gyulassy,
\newblock Phys.Rev.Lett. {\bf 68}, 1480 (1992).

\bibitem{Qin:2015srf}
G.-Y. Qin and X.-N. Wang,
\newblock Int. J. Mod. Phys. {\bf E24}, 1530014 (2015), arXiv:1511.00790.

\bibitem{Baier:1996sk}
R.~Baier, Y.~L. Dokshitzer, A.~H. Mueller, S.~Peigne, and D.~Schiff,
\newblock Nucl.Phys. {\bf B484}, 265 (1997), arXiv:hep-ph/9608322.

\bibitem{Majumder:2007zh}
A.~Majumder, B.~Muller, and X.-N. Wang,
\newblock Phys.Rev.Lett. {\bf 99}, 192301 (2007), arXiv:hep-ph/0703082.

\bibitem{Khachatryan:2016odn}
CMS, V.~Khachatryan {\em et~al.},
\newblock JHEP {\bf 04}, 039 (2017), arXiv:1611.01664.

\bibitem{Acharya:2018qsh}
ALICE, S.~Acharya {\em et~al.},
\newblock JHEP {\bf 11}, 013 (2018), arXiv:1802.09145.

\bibitem{Aad:2015wga}
ATLAS, G.~Aad {\em et~al.},
\newblock JHEP {\bf 09}, 050 (2015), arXiv:1504.04337.

\bibitem{Burke:2013yra}
JET, K.~M. Burke {\em et~al.},
\newblock Phys. Rev. {\bf C90}, 014909 (2014), arXiv:1312.5003.

\bibitem{Xu:2014tda}
J.~Xu, J.~Liao, and M.~Gyulassy,
\newblock Chin. Phys. Lett. {\bf 32}, 092501 (2015), arXiv:1411.3673.

\bibitem{Chien:2015vja}
Y.-T. Chien, A.~Emerman, Z.-B. Kang, G.~Ovanesyan, and I.~Vitev,
\newblock Phys. Rev. {\bf D93}, 074030 (2016), arXiv:1509.02936.

\bibitem{Andres:2016iys}
C.~Andrés, N.~Armesto, M.~Luzum, C.~A. Salgado, and P.~Zurita,
\newblock Eur. Phys. J. {\bf C76}, 475 (2016), arXiv:1606.04837.

\bibitem{Cao:2017hhk}
S.~Cao, T.~Luo, G.-Y. Qin, and X.-N. Wang,
\newblock Phys. Lett. {\bf B777}, 255 (2018), arXiv:1703.00822.

\bibitem{Zigic:2018ovr}
D.~Zigic, I.~Salom, J.~Auvinen, M.~Djordjevic, and M.~Djordjevic,
\newblock Phys. Lett. {\bf B791}, 236 (2019), arXiv:1805.04786.

\bibitem{Xing:2019xae}
W.-J. Xing, S.~Cao, G.-Y. Qin, and H.~Xing,
\newblock (2019), arXiv:1906.00413.

\bibitem{Qin:2009bk}
G.-Y. Qin, J.~Ruppert, C.~Gale, S.~Jeon, and G.~D. Moore,
\newblock Phys.Rev. {\bf C80}, 054909 (2009), arXiv:0906.3280.

\bibitem{Chen:2016vem}
L.~Chen, G.-Y. Qin, S.-Y. Wei, B.-W. Xiao, and H.-Z. Zhang,
\newblock Phys. Lett. {\bf B773}, 672 (2017), arXiv:1607.01932.

\bibitem{Chen:2016cof}
L.~Chen, G.-Y. Qin, S.-Y. Wei, B.-W. Xiao, and H.-Z. Zhang,
\newblock Phys. Lett. {\bf B782}, 773 (2018), arXiv:1612.04202.

\bibitem{Chen:2017zte}
W.~Chen, S.~Cao, T.~Luo, L.-G. Pang, and X.-N. Wang,
\newblock Phys. Lett. {\bf B777}, 86 (2018), arXiv:1704.03648.

\bibitem{Luo:2018pto}
T.~Luo, S.~Cao, Y.~He, and X.-N. Wang,
\newblock Phys. Lett. {\bf B782}, 707 (2018), arXiv:1803.06785.

\bibitem{Zhang:2018urd}
S.-L. Zhang, T.~Luo, X.-N. Wang, and B.-W. Zhang,
\newblock Phys. Rev. {\bf C98}, 021901 (2018), arXiv:1804.11041.

\bibitem{Kang:2018wrs}
Z.-B. Kang, J.~Reiten, I.~Vitev, and B.~Yoon,
\newblock Phys. Rev. {\bf D99}, 034006 (2019), arXiv:1810.10007.

\bibitem{Aad:2010bu}
Atlas Collaboration, G.~Aad {\em et~al.},
\newblock Phys.Rev.Lett. {\bf 105}, 252303 (2010), arXiv:1011.6182.

\bibitem{Chatrchyan:2011sx}
CMS, S.~Chatrchyan {\em et~al.},
\newblock Phys. Rev. {\bf C84}, 024906 (2011), arXiv:1102.1957.

\bibitem{Chatrchyan:2012gt}
CMS, S.~Chatrchyan {\em et~al.},
\newblock Phys. Lett. {\bf B718}, 773 (2013), arXiv:1205.0206.

\bibitem{Aad:2014bxa}
ATLAS, G.~Aad {\em et~al.},
\newblock Phys. Rev. Lett. {\bf 114}, 072302 (2015), arXiv:1411.2357.

\bibitem{Chatrchyan:2012gw}
CMS Collaboration, S.~Chatrchyan {\em et~al.},
\newblock JHEP {\bf 1210}, 087 (2012), arXiv:1205.5872.

\bibitem{Chatrchyan:2013kwa}
CMS Collaboration, S.~Chatrchyan {\em et~al.},
\newblock Phys.Lett. {\bf B730}, 243 (2014), arXiv:1310.0878.

\bibitem{Aad:2014wha}
ATLAS, G.~Aad {\em et~al.},
\newblock Phys. Lett. {\bf B739}, 320 (2014), arXiv:1406.2979.

\bibitem{Vitev:2009rd}
I.~Vitev and B.-W. Zhang,
\newblock Phys. Rev. Lett. {\bf 104}, 132001 (2010), arXiv:0910.1090.

\bibitem{Qin:2010mn}
G.-Y. Qin and B.~Muller,
\newblock Phys. Rev. Lett. {\bf 106}, 162302 (2011), arXiv:1012.5280,
\newblock [Erratum: Phys. Rev. Lett.108,189904(2012)].

\bibitem{CasalderreySolana:2010eh}
J.~Casalderrey-Solana, J.~G. Milhano, and U.~A. Wiedemann,
\newblock J. Phys. {\bf G38}, 035006 (2011), arXiv:1012.0745.

\bibitem{Lokhtin:2011qq}
I.~Lokhtin, A.~Belyaev, and A.~Snigirev,
\newblock Eur.Phys.J. {\bf C71}, 1650 (2011), arXiv:1103.1853.

\bibitem{Young:2011qx}
C.~Young, B.~Schenke, S.~Jeon, and C.~Gale,
\newblock Phys.Rev. {\bf C84}, 024907 (2011), arXiv:1103.5769.

\bibitem{He:2011pd}
Y.~He, I.~Vitev, and B.-W. Zhang,
\newblock Phys.Lett. {\bf B713}, 224 (2012), arXiv:1105.2566.

\bibitem{MehtarTani:2011tz}
Y.~Mehtar-Tani, C.~Salgado, and K.~Tywoniuk,
\newblock Phys.Lett. {\bf B707}, 156 (2012), arXiv:1102.4317.

\bibitem{Renk:2012cx}
T.~Renk,
\newblock Phys.Rev. {\bf C85}, 064908 (2012), arXiv:1202.4579.

\bibitem{Zapp:2012ak}
K.~C. Zapp, F.~Krauss, and U.~A. Wiedemann,
\newblock JHEP {\bf 03}, 080 (2013), arXiv:1212.1599.

\bibitem{Apolinario:2012cg}
L.~Apolinario, N.~Armesto, and L.~Cunqueiro,
\newblock JHEP {\bf 02}, 022 (2013), arXiv:1211.1161.

\bibitem{Dai:2012am}
W.~Dai, I.~Vitev, and B.-W. Zhang,
\newblock Phys. Rev. Lett. {\bf 110}, 142001 (2013), arXiv:1207.5177.

\bibitem{Qin:2012gp}
G.-Y. Qin,
\newblock Eur.Phys.J. {\bf C74}, 2959 (2014), arXiv:1210.6610.

\bibitem{CasalderreySolana:2012ef}
J.~Casalderrey-Solana, Y.~Mehtar-Tani, C.~A. Salgado, and K.~Tywoniuk,
\newblock Phys.Lett. {\bf B725}, 357 (2013), arXiv:1210.7765.

\bibitem{Majumder:2013re}
A.~Majumder,
\newblock Phys.Rev. {\bf C88}, 014909 (2013), arXiv:1301.5323.

\bibitem{Ma:2013pha}
G.-L. Ma,
\newblock Phys. Rev. {\bf C87}, 064901 (2013), arXiv:1304.2841.

\bibitem{Senzel:2013dta}
F.~Senzel, O.~Fochler, J.~Uphoff, Z.~Xu, and C.~Greiner,
\newblock J. Phys. {\bf G42}, 115104 (2015), arXiv:1309.1657.

\bibitem{Blaizot:2013hx}
J.-P. Blaizot, E.~Iancu, and Y.~Mehtar-Tani,
\newblock Phys.Rev.Lett. {\bf 111}, 052001 (2013), arXiv:1301.6102.

\bibitem{Wang:2013cia}
X.-N. Wang and Y.~Zhu,
\newblock Phys. Rev. Lett. {\bf 111}, 062301 (2013), arXiv:1302.5874.

\bibitem{Fister:2014zxa}
L.~Fister and E.~Iancu,
\newblock JHEP {\bf 03}, 082 (2015), arXiv:1409.2010.

\bibitem{Casalderrey-Solana:2014bpa}
J.~Casalderrey-Solana, D.~C. Gulhan, J.~G. Milhano, D.~Pablos, and
  K.~Rajagopal,
\newblock JHEP {\bf 10}, 019 (2014), arXiv:1405.3864,
\newblock [Erratum: JHEP09,175(2015)].

\bibitem{He:2015pra}
Y.~He, T.~Luo, X.-N. Wang, and Y.~Zhu,
\newblock Phys. Rev. {\bf C91}, 054908 (2015), arXiv:1503.03313,
\newblock [Erratum: Phys. Rev.C97,no.1,019902(2018)].

\bibitem{Chien:2015hda}
Y.-T. Chien and I.~Vitev,
\newblock JHEP {\bf 05}, 023 (2016), arXiv:1509.07257.

\bibitem{Milhano:2015mng}
J.~Milhano and K.~C. Zapp,
\newblock Eur. Phys. J. {\bf C76}, 288 (2016), arXiv:1512.08107.

\bibitem{Chang:2016gjp}
N.-B. Chang and G.-Y. Qin,
\newblock Phys. Rev. {\bf C94}, 024902 (2016), arXiv:1603.01920.

\bibitem{Tachibana:2017syd}
Y.~Tachibana, N.-B. Chang, and G.-Y. Qin,
\newblock Phys. Rev. {\bf C95}, 044909 (2017), arXiv:1701.07951.

\bibitem{Chien:2016led}
Y.-T. Chien and I.~Vitev,
\newblock Phys. Rev. Lett. {\bf 119}, 112301 (2017), arXiv:1608.07283.

\bibitem{Chang:2017gkt}
N.-B. Chang, S.~Cao, and G.-Y. Qin,
\newblock Phys. Lett. {\bf B781}, 423 (2018), arXiv:1707.03767.

\bibitem{Cao:2017zih}
JETSCAPE, S.~Cao {\em et~al.},
\newblock Phys. Rev. {\bf C96}, 024909 (2017), arXiv:1705.00050.

\bibitem{Cao:2017qpx}
S.~Cao and A.~Majumder,
\newblock (2017), arXiv:1712.10055.

\bibitem{He:2018xjv}
Y.~He {\em et~al.},
\newblock Phys. Rev. {\bf C99}, 054911 (2019), arXiv:1809.02525.

\bibitem{Wicks:2005gt}
S.~Wicks, W.~Horowitz, M.~Djordjevic, and M.~Gyulassy,
\newblock Nucl. Phys. {\bf A784}, 426 (2007), arXiv:nucl-th/0512076.

\bibitem{Qin:2007rn}
G.-Y. Qin {\em et~al.},
\newblock Phys. Rev. Lett. {\bf 100}, 072301 (2008), arXiv:0710.0605.

\bibitem{Schenke:2009ik}
B.~Schenke, C.~Gale, and G.-Y. Qin,
\newblock Phys. Rev. {\bf C79}, 054908 (2009), arXiv:0901.3498.

\bibitem{Cao:2013ita}
S.~Cao, G.-Y. Qin, and S.~A. Bass,
\newblock Phys.Rev. {\bf C88}, 044907 (2013), arXiv:1308.0617.

\bibitem{Cao:2015hia}
S.~Cao, G.-Y. Qin, and S.~A. Bass,
\newblock Phys. Rev. {\bf C92}, 024907 (2015), arXiv:1505.01413.

\bibitem{KunnawalkamElayavalli:2017hxo}
R.~Kunnawalkam~Elayavalli and K.~C. Zapp,
\newblock JHEP {\bf 07}, 141 (2017), arXiv:1707.01539.

\bibitem{Sirunyan:2018ncy}
CMS, A.~M. Sirunyan {\em et~al.},
\newblock Phys. Rev. Lett. {\bf 122}, 152001 (2019), arXiv:1809.08602.

\bibitem{Wang:2001ifa}
X.-N. Wang and X.-f. Guo,
\newblock Nucl. Phys. {\bf A696}, 788 (2001), arXiv:hep-ph/0102230.

\bibitem{Majumder:2009ge}
A.~Majumder,
\newblock Phys. Rev. {\bf D85}, 014023 (2012), arXiv:0912.2987.

\bibitem{Chen:2010te}
X.-F. Chen, C.~Greiner, E.~Wang, X.-N. Wang, and Z.~Xu,
\newblock Phys. Rev. {\bf C81}, 064908 (2010), arXiv:1002.1165.

\bibitem{Baier:2006pt}
R.~Baier, A.~H. Mueller, and D.~Schiff,
\newblock Phys. Lett. {\bf B649}, 147 (2007), arXiv:nucl-th/0612068.

\bibitem{Moore:2004tg}
G.~D. Moore and D.~Teaney,
\newblock Phys.Rev. {\bf C71}, 064904 (2005), arXiv:hep-ph/0412346.

\bibitem{Qin:2009gw}
G.-Y. Qin and A.~Majumder,
\newblock Phys.Rev.Lett. {\bf 105}, 262301 (2010), arXiv:0910.3016.

\bibitem{Song:2007fn}
H.~Song and U.~W. Heinz,
\newblock Phys.Lett. {\bf B658}, 279 (2008), arXiv:0709.0742.

\bibitem{Schenke:2010rr}
B.~Schenke, S.~Jeon, and C.~Gale,
\newblock Phys.Rev.Lett. {\bf 106}, 042301 (2011), arXiv:1009.3244.

\bibitem{Schenke:2011tv}
B.~Schenke, S.~Jeon, and C.~Gale,
\newblock Phys. Lett. {\bf B702}, 59 (2011), arXiv:1102.0575.

\bibitem{Song:2011qa}
H.~Song, S.~A. Bass, and U.~Heinz,
\newblock Phys. Rev. {\bf C83}, 054912 (2011), arXiv:1103.2380,
\newblock [Erratum: Phys. Rev.C87,no.1,019902(2013)].

\bibitem{Petersen:2011sb}
H.~Petersen,
\newblock Phys. Rev. {\bf C84}, 034912 (2011), arXiv:1105.1766.

\bibitem{Shen:2011eg}
C.~Shen, U.~Heinz, P.~Huovinen, and H.~Song,
\newblock Phys. Rev. {\bf C84}, 044903 (2011), arXiv:1105.3226.

\bibitem{Yan:2017rku}
L.~Yan, S.~Jeon, and C.~Gale,
\newblock Phys. Rev. {\bf C97}, 034914 (2018), arXiv:1707.09519.

\bibitem{Hirano:2005xf}
T.~Hirano, U.~W. Heinz, D.~Kharzeev, R.~Lacey, and Y.~Nara,
\newblock Phys. Lett. {\bf B636}, 299 (2006), arXiv:nucl-th/0511046.

\bibitem{Adam:2015kda}
ALICE, J.~Adam {\em et~al.},
\newblock Phys. Lett. {\bf B754}, 373 (2016), arXiv:1509.07299.

\bibitem{Adam:2016ddh}
ALICE, J.~Adam {\em et~al.},
\newblock Phys. Lett. {\bf B772}, 567 (2017), arXiv:1612.08966.

\bibitem{Noronha-Hostler:2016eow}
J.~Noronha-Hostler, B.~Betz, J.~Noronha, and M.~Gyulassy,
\newblock Phys. Rev. Lett. {\bf 116}, 252301 (2016), arXiv:1602.03788.

\bibitem{Borsanyi:2013bia}
S.~Borsanyi {\em et~al.},
\newblock Phys. Lett. {\bf B730}, 99 (2014), arXiv:1309.5258.

\bibitem{Cooper:1974mv}
F.~Cooper and G.~Frye,
\newblock Phys.Rev. {\bf D10}, 186 (1974).

\bibitem{Sirunyan:2018jqr}
CMS, A.~M. Sirunyan {\em et~al.},
\newblock JHEP {\bf 05}, 006 (2018), arXiv:1803.00042.

\bibitem{Miller:2007ri}
M.~L. Miller, K.~Reygers, S.~J. Sanders, and P.~Steinberg,
\newblock Ann. Rev. Nucl. Part. Sci. {\bf 57}, 205 (2007),
  arXiv:nucl-ex/0701025.

\bibitem{Qin:2010pf}
G.-Y. Qin, H.~Petersen, S.~A. Bass, and B.~Muller,
\newblock Phys.Rev. {\bf C82}, 064903 (2010), arXiv:1009.1847.

\bibitem{Sjostrand:2007gs}
T.~Sjostrand, S.~Mrenna, and P.~Z. Skands,
\newblock Comput.Phys.Commun. {\bf 178}, 852 (2008), arXiv:0710.3820.

\bibitem{Cacciari:2011ma}
M.~Cacciari, G.~P. Salam, and G.~Soyez,
\newblock Eur.Phys.J. {\bf C72}, 1896 (2012), arXiv:1111.6097.

\bibitem{Sjostrand:2006za}
T.~Sjostrand, S.~Mrenna, and P.~Z. Skands,
\newblock JHEP {\bf 05}, 026 (2006), arXiv:hep-ph/0603175.

\bibitem{Hosokawa:2019odr}
ALICE, R.~Hosokawa,
\newblock Nucl. Phys. {\bf A982}, 639 (2019).

\bibitem{Aad:2012vca}
ATLAS, G.~Aad {\em et~al.},
\newblock Phys. Lett. {\bf B719}, 220 (2013), arXiv:1208.1967.

\bibitem{Khachatryan:2016jfl}
CMS, V.~Khachatryan {\em et~al.},
\newblock Phys. Rev. {\bf C96}, 015202 (2017), arXiv:1609.05383.

\bibitem{Wang:1996yh}
X.-N. Wang, Z.~Huang, and I.~Sarcevic,
\newblock Phys.Rev.Lett. {\bf 77}, 231 (1996), arXiv:hep-ph/9605213.

\bibitem{Zhang:2009rn}
H.~Zhang, J.~F. Owens, E.~Wang, and X.-N. Wang,
\newblock Phys. Rev. Lett. {\bf 103}, 032302 (2009), arXiv:0902.4000.

\bibitem{Sirunyan:2017qhf}
CMS, A.~M. Sirunyan {\em et~al.},
\newblock Phys. Lett. {\bf B785}, 14 (2018), arXiv:1711.09738.
\end{thebibliography}

\end{document}